\documentclass[twocolumn,showpacs,aps,prb]{revtex4}
\usepackage{CJK}
\usepackage{graphicx}% Include figure files
\usepackage{epsfig}% Include figure files
\usepackage{latexsym}
\usepackage{amsmath}
\usepackage{wasysym}
\usepackage{nicefrac}
\newcommand{\sumnear}{\mathop{\sum}_{\langle i j \rangle}}
\begin{document}
\title{Chiral edge states and fractional charge separation in interacting
bosons on a Kagome lattice\\~\\}
\author{Xue-Feng Zhang and Sebastian Eggert}
\affiliation{Physics Department and Research Center OPTIMAS, University of
Kaiserslautern, 67663 Kaiserslautern, Germany}
\begin{abstract}
We consider the extended hard-core Bose-Hubbard model 
on a Kagome lattice with boundary conditions on two edges.   
We find that the sharp edges lift the degeneracy and 
freeze the system into a striped order at $\nicefrac{1}{3}$ and $\nicefrac{2}{3}$
filling for zero hopping.
At small hopping strengths, holes spontaneously appear and 
separate into fractional charges which move to
 the edges of the system. This leads to a novel edge liquid phase, 
which is characterized by fractional charges near the edges and a finite edge
compressibility but no superfluid density.
The compressibility is due to excitations on the edge which display a 
chrial symmetry breaking that is reminiscent of the quantum Hall effect and 
topological insulators.  
Large scale Monte Carlo simulations confirm the analytical considerations.
\end{abstract}
\maketitle

 Frustrated systems are a rich playground in the search for
new exotic phases and excitations, such as spin liquid
phases,\cite{spinliquid1,spinliquid2} Dirac strings in a
spin-ice,\cite{spinice1,spinice2} and fractional charges in Kagome
lattice antiferromagnets.\cite{pollman1} Recently, progress of
indirect observations of fractional excitations has been
made,\cite{frac_observ} but their detection remains far from
trivial since the controlled excitation and separation of
fractional charges is difficult. We now show that by introducing
sharp edges on two sides of a Kagome lattice with interacting
bosons, fractional charges appear spontaneously and are located
close to the separate edges depending on their chirality.   These
chiral edge states are reminiscent of phenomena in quantum hall
physics\cite{QHE} and topological insulators,\cite{TI} but on the
Kagome lattice the topology and chirality is completely controlled
by the design of the edges.  Moreover, the appearance of the
chiral fractional charges with a deconfined interaction gives rise
to a new compressible quantum edge phase.

%fractional excitation
Exotic excitations in frustrated systems such as fractional
charges and monopoles can often be understood from rather
straight-forward geometrical arguments.\cite{balents-rev} They
appear as local defects in the form of a rearrangement of real
charges on the nontrivial
background.\cite{spinice2,a_kagome1,pollman1,spinliquid3} When two
(or more) fractional charges separate they may be connected by a
{\it string} of a slightly disturbed quantum ground state, which
normally acts confining. This poses a generic problem for the controlled
excitation and observation of the fractional charges: the confining string keeps
them close together so that a separate observation becomes
impossible, analogously to the difficulty of detecting individual
quarks. If the string can be tuned to become deconfining, then the
system typically undergoes a quantum phase transitions to a complicated sea
of closed strings with entirely different properties.

%introduce the model and the effect of the frustration
We now consider a system of interacting bosons on a Kagome
lattice where boundary conditions play an important role due to the huge degeneracy,
and deconfining strings appear, which allow a controlled separation of the fractional 
charges.
The model is defined by hard core bosons on a Kagome lattice
with nearest neighbor repulsion $V$ and chemical potential $\mu$
\begin{eqnarray}
H&=&-t\sumnear(b_{i}^{\dag}b_{j}^{\phantom{\dag}}+b_{j}^{\dag}b_{i}^{\phantom{\dag}})+V\sumnear
n_{i}n_{j} -\mu\mathop{\sum}_in_i. \label{BH}
\end{eqnarray}
where $t$ is the hopping between boson creation and annihilation
operators $b_{i}^{\dag}$ and  $b_j^{\phantom{\dag}}$ on nearest
neighbor sites $\langle i,j\rangle$. 
The Kagome lattice is special since it
allows a macroscopic degeneracy even for commensurate fillings of
$\nicefrac{1}{3}$ and $\nicefrac{2}{3}$. In particular, without hopping the model in
Eq.~(\ref{BH}) corresponds to the Ising model with an entropy per
site\cite{entropy} of $S=0.108$ at filling $\nicefrac{1}{3}$, since all
configurations with one particle per triangle are ground states
i.e.~there is a ``triangle rule" analogous to the ``ice rule" in
the spin ice.\cite{spinice2} In contrast, the Ising model on the
triangular lattice, which is also frustrated, does not show a
macroscopic degeneracy at commensurate filling. Finite hopping
$t>0$ lifts this degeneracy which leads to an ordered state with a
finite structure factor. At even larger hopping a phase transition
to a superfluid phase is observed, which is believed to be weakly
first order or possibly second order.\cite{wessel} At negative
hopping $t=-\nicefrac{V}{2}$ the model corresponds to the Heisenberg model
which is at half filling one of the most promising candidate for a
spin liquid.\cite{spinliquid2}

\begin{figure}
\includegraphics[width=0.5\textwidth]{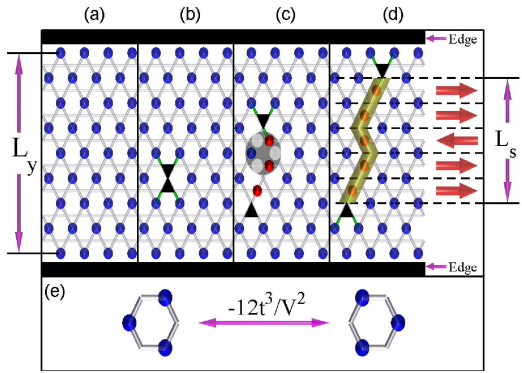}%
\caption{ (a) Striped solid phase induced by edges. (b) One hole
corresponds to two fractional charges (black triangles). (c)
Hopping of real particles (blue dots) separates the fractional
charges and resonant hexagons appear along the connecting
string (red dots). (d) The spin (red arrows) representation of the
string (yellow path). (e) Resonant hexagons can be flipped by
a third order hopping process. \label{fig1}}
\end{figure}

%what's the fractional charges and hard to find it
Fractional charges on the Kagome lattice which are connected by
strings have been extensively discussed\cite{pollman1} for the fermionic model in
Eq.~(\ref{BH}) at filling $\nicefrac{1}{3}$.  Basically the
fractional charges correspond to one empty triangle, which have
two possible chiralities:  up or down. Since one up and one down
triangle together correspond to a single missing charge (see
Fig.~\ref{fig1}b), it is clear that a single empty triangle has
negative fractional bosonic charge -$\nicefrac{1}{2}$. However, so far there is no
proposal on how to directly observe these exotic excitations.

%the t=0 limit
As we will show here, it is possible to make a controlled
separation of fractional charges by introducing sharp edges in the
model of Eq.~(\ref{BH}) on two sides of the Kagome lattice, which
also gives rise to a new boundary-induced quantum edge phase.
Interestingly, such edges as depicted in Fig.~\ref{fig1} have a
macroscopic effect in the Kagome lattice, since the edges fix the
bosons at the upper corners of the triangles. Therefore the order
becomes frozen for $t=0$ and there is a unique ground state of
bosons in stripes parallel to the edges as shown in
Fig.~\ref{fig1}a. In other words, the macroscopic entropy is lost
in this ``striped solid phase", even when the edges are very far
apart.

%excitation on this system
In a system with such edges the role of finite but small hopping
$t$ is now dramatically reversed:  Instead of lifting the
degeneracy, the hopping now facilitates fluctuations, which turn
out to be nothing else but strings between fractional charges. As
depicted in Fig.~\ref{fig1} the separation of two fractional
charges facilitates a string with configurations of triple
occupied hexagons. At finite hopping the triple occupied hexagons
are preferred since they lower the energy by $-12t^3/V^2$ via a
resonant third order hopping process\cite{wessel} as shown in
Fig.~\ref{fig1}e.

%why is it interesting
At this point two interesting observations can be made:  First of
all there is a new competition between a frozen ground state due
to the edges without any resonant hexagons (striped solid phase)
on the one hand and possible fluctuating strings on the other
hand. Secondly, closed strings cannot appear by a local
rearrangement of charges without violating the triangle rule on
the frozen configuration shown in Fig.~\ref{fig1}a. Indeed it is
straight-forward to convince oneself that fluctuating strings with
triple occupied hexagons can only occur if either the triangle
rule is violated or holes are introduced into the
system.  For the case $\mu<V$ we therefore expect that holes
may appear spontaneously which separate into two fractional charges since the
connecting strings allow additional resonant hexagons and
therefore are {\it deconfining}.  However, this phase is not the
superfluid phase since the fractional charges are not free. Indeed
we will see that the fractional charges only appear as edge
excitations, so we will call this phase the ``edge liquid phase".

%explain the process in detail
In order to establish the new edge phase quantitatively, it is
important to understand  the process of fractional charge
separation in detail as illustrated in Fig.~\ref{fig1}.  In
Fig.~\ref{fig1}a the frozen configuration in the striped order is
shown, which is the unique ground state for $t=0$. If a charge is
removed (Fig.~\ref{fig1}b) two fractional charges can separate for
$t\neq0$, namely one up and one down triangle (Fig.~\ref{fig1}c)
which are connected by a {string} (red dots). Because of this
string, the down triangle can never move underneath the up
triangle. Since there are now resonant configurations (hexagons
with three bosons) along the string,  it is deconfining and the
energy can be lowered by pushing the fractional charges to the
upper and lower edges respectively. When the energy gain from a
fluctuating string exceeds the energy cost $\mu$ of removing a
charge, it can be expected that fractional charges appear
spontaneously at the edges, leading to a new quantum edge phase
which is compressible.

%Estimating the energy of string: effective model
The energy of a string can in turn be quantitatively estimated by
realizing that in fact the resonances on each hexagon correspond
to fluctuations of the string as shown in Fig.~\ref{fig1}d. Here
the path of the string is mapped to up and down spins depending on
if the string is to the right or left of a resonant hexagon.  A
flip can occur when the neighboring hexagon has opposite spin,
leading to an ${\rm xy}-$type model with effective exchange of
$-12t^3/V^2$ along the length of the string $L_s \leq L_y$
\begin{eqnarray}
H_{\rm xy}&=&
- \frac{12t^3}{V^2}\mathop{\sum}_{i=1}^{L_{s}}(S_i^+S_{i+1}^-+h.c.).\label{xy}
\end{eqnarray}
For a given length $L_s$ the ground state energy of this model is given by
\begin{eqnarray}
E_{\rm xy}(L_s)= \frac{12t^3}{V^2}\left(1-\cot\frac{\pi}{2(L_s+1)} \right)
\approx -\frac{24t^3}{\pi V^2}{L_s},
\end{eqnarray}
which means that there can be a substantial energy gain by
maximizing the length $L_s$. However, this is only part of the
story since the ends of the string can move freely up to the total
systems size $L_y \geq L_s$, due to the hopping of the fractional
charges.  This hopping gives an additional kinetic energy of order
$t$ and independent of length, but for large systems sizes the
leading effect of the ${\rm xy}-$model in Eq.~(\ref{xy}) is to
maximize the length of the string $L_s$, i.e.~push the fractional
charges apart towards the edges by an effective linear repulsive
potential between the fractional charges. The interplay of
maximizing $L_s$ and the kinetic energy of the fractional charges
leads to a characteristic maximum of fractional charges near the
edges, which resembles the solution of a particle in a linear
potential (Airy function). At the top edge the fractional charges
appear as down-triangles only and at the bottom edge they appear
as up-triangles. These edge excitations therefore posses a
definite chirality, which relates the pseudo-spin of the
fractional charge (up or down) to the direction given by the edge.

\begin{figure}
\includegraphics[width=0.5\textwidth]{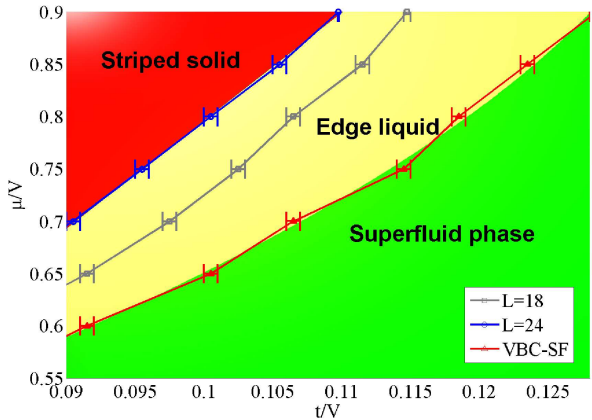}%
\caption{The phase diagram of the Bose-Hubbard model with
cylindrical boundary conditions for $L_y=18$ and $24$ from Monte
Carlo simulations at temperature $T=0.005V$. The phase transition
line to the superfluid phase was determined on lattices up to
$L_y=24$ with periodic boundary conditions. \label{phase}}
\end{figure}

%Show the algorithm and the physical variables
We now turn to quantum Monte Carlo simulations of the hard-core
boson model in Eq.~(\ref{BH}) on the Kagome lattice in order to
numerically analyze the existence of the edge excitations. We use
the stochastic cluster series expansion\cite{sse,clustersse} with
parallel tempering\cite{para1,para2,imp2} on system sizes up to
$1752$ sites ($L_y=L_x=24$) with cylindrical boundary conditions.
In Fig.~\ref{phase} we show a small section of the phase
diagram\cite{wessel} near the transition to the $\nicefrac{1}{3}$ solid phase
for cylindrical boundary conditions.  As predicted, the striped
solid order is destroyed by the spontaneous appearance of
fractional charges as the hopping is increased.  The value of the
critical hopping is lower for larger $L_y$, which is in agreement
with the argument that the energy of the string is proportional to
$L_s$. In fact, the effective model above can also be used to
determine the critical hopping, which gives a compatible estimate
at about 5\% larger hopping. In the thermodynamic limit, the
energy of strings will always be lower than the chemical potential
and the edge liquid phase exists for all hopping strengths.

\begin{figure}
\includegraphics[width=0.5\textwidth]{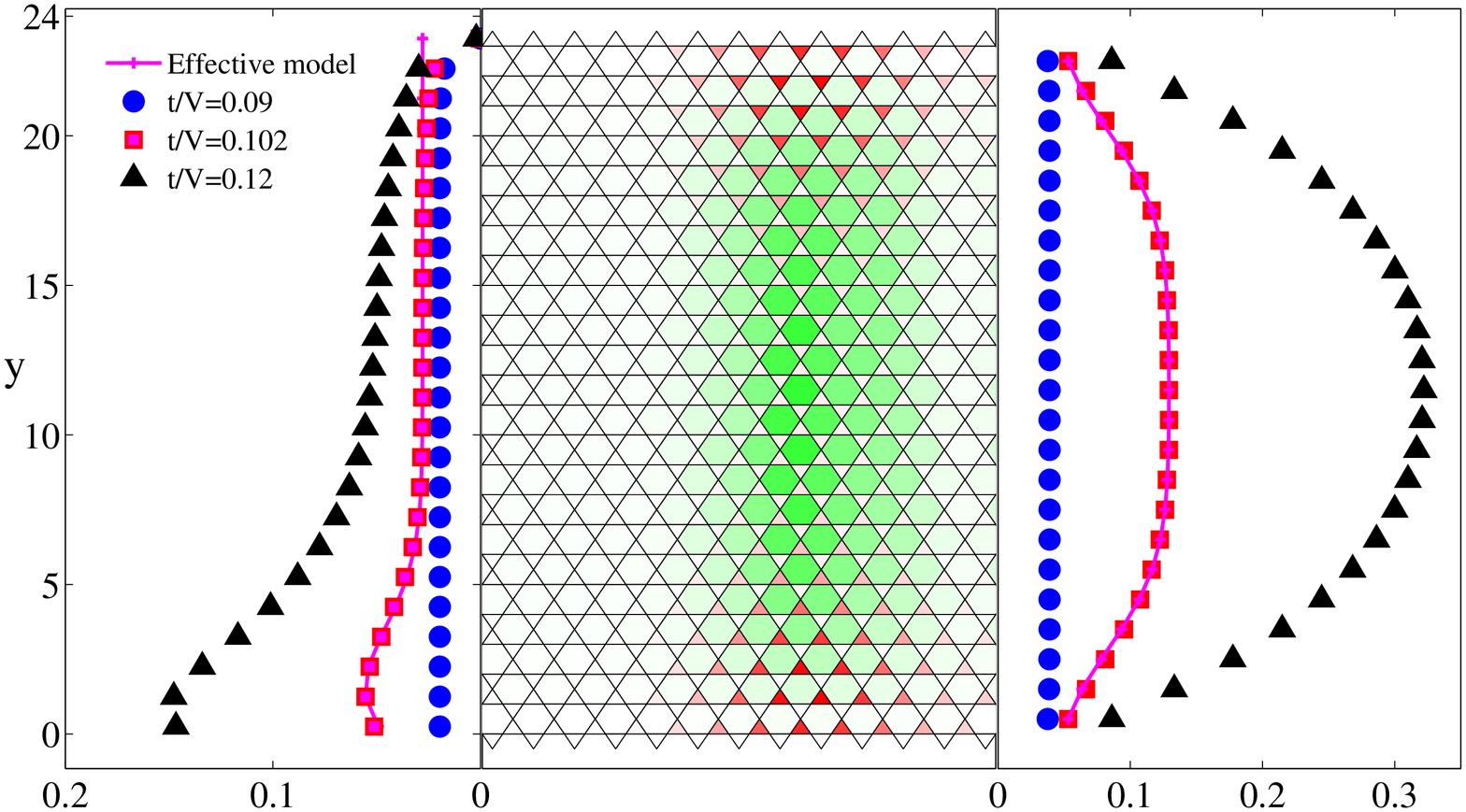}%
\caption{A snapshot of a typical configuration during a Monte
Carlo run at $\mu/V=0.8$, $t/V=0.102$ and $\beta=200$ with
$L_x=L_y=24$ is shown in the middle panel. Red 
indicates a higher density of empty triangles or fractional charges and green
represents a higher density of hexagons in a resonant configuration. The
density distribution of up-triangle fractional charges is shown in
the left panel for the striped solid ($t=0.09V$), edge liquid
($t=0.102V$) and superfluid ($t=0.12$) phases.  The right panel
shows the corresponding densities of resonant hexagons.
\label{fig3}}
\end{figure}

In the middle panel of Fig.~\ref{fig3}, a typical snapshot of a
configuration during a Monte Carlo simulations for $t=0.102 V$ in
the edge liquid phase is shown.  A string is
clearly visible by the shaded resonant hexagons (green) and the
shaded  up and down triangles (red) which indicate fractional
charges near the edges. The distribution of up-triangle fractional
charges is shown in the left panel of Fig.~\ref{fig3} with a
characteristic maximum at the lower edge, which agrees well with
the results from the effective model discussed in Eq.~(\ref{xy}).
In the striped solid phase ($t=0.09V$) no fractional charges are
found (except for virtual excitations) and in the superfluid phase
($t=1.2V$) there are fractional charges in the entire sample, but
the maximum near the edge remains. The right panel of
Fig.~\ref{fig3} shows the change in resonant hexagon density in
the different phases.

\begin{figure}
\includegraphics[width=0.5\textwidth]{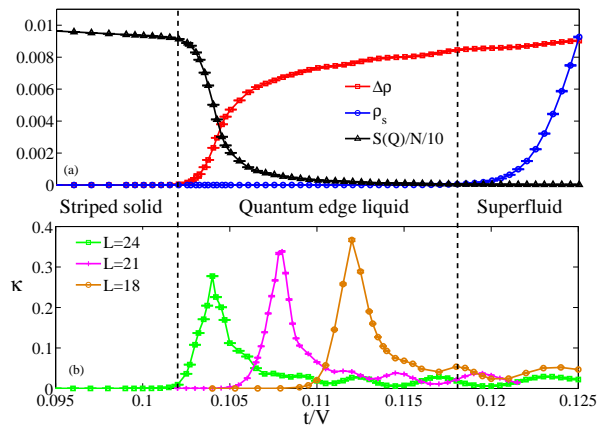}%
\caption{(a) The change of hole density $\Delta\rho$ relative to the striped solid phase, 
the structure factor with $Q=(2 \pi,0)$ and the superfluid density
$\rho_s$ at $\mu/V=0.8$ and $\beta=200$ with $L_{\rm x}=L_{\rm
y}=24$ in different phases. (b) The compressibility at
$\mu/V=0.8$ for different sizes. \label{fig4}}
\end{figure}

The relevant order parameters in the different phases are shown in
Fig.~\ref{fig4}. The spontaneous appearance of additional charges
at a critical hopping is clearly seen by the change of total hole
density $\Delta \rho$ (relative to the striped order phase). The
structure factor $S(\textbf{Q})/N=\langle |\mathop{\sum}_{k=1}^N
n_k e^{\emph{\textbf{i}} \textbf{Q}
\cdot\textbf{r}_k}|^2\rangle/N^2$ at $\textbf{Q}=(2 \pi,0)$ is an
indication of the striped order, which drops sharply when the
fractional charges appear in the edge liquid phase, but remains
finite. At still larger hopping the system enters the superfluid
phase, which has a spontaneously broken U(1) symmetry
(off-diagonal order) and a finite superfluid density
($\rho_s=\langle W^2/2\beta t\rangle$ in terms of the winding
number $W$ in Monte Carlo simulations\cite{winding}). The
compressibility per site $\kappa$ is maybe the most interesting
order parameter, since it is zero in the striped order phase and
then has a clear maximum close to the phase transition into the
edge liquid phase. The bottom panel of Fig.~\ref{fig4} illustrates
how this maximum (i.e.~the phase transition) moves to lower values
of $t$ as the system size increases.  At the same time, the
compressibility per site {\it decreases} proportional to $1/L_y$,
which is a clear indication that the compressibility is due to an
``edge compressibility".  Indeed according to the considerations
above the bulk sites do not contribute to the compressibility
since there are no fractional charges (see Fig.~\ref{fig3}). The
phase transition is basically characterized by the filling of
interacting strings which makes it second order. The system is
most compressible at the phase transition, but additional strings
are harder and harder to insert which leads to a characteristic
maximum. The oscillations of the compressibility as a function of
hopping $t$ are a finite size effect, where the addition of an
extra string leads to regular blockades before the next string can
be added, analogously to the quantized charging energy of a small
capacitor (Coulomb blockade).

In summary, the results show that edges have a dramatic effect for
hard core bosons on a Kagome lattice and lead to a  new edge
liquid phase, which is characterized by a finite edge
compressibility but no superfluid density. Quantum Monte Carlo
simulations and analytical arguments demonstrate that fractional
excitations of up- and down-triangles can be separated in space
and are localized close to the lower and upper edges respectively.
The fractional charges are connected by quantum strings of
resonant configurations, which act deconfining. Edge states with a
finite compressibility and the locking of an internal quantum
number with the direction of the edge (e.g.~the chirality of the
spin-momentum locking) are also famous characteristics of
topological insulators and the quantum Hall effect.  However, for
the hard-core boson system studied here the non-trivial topology
is not a property of the bulk, but is instead controlled by the
design of the sharp edges.  In fact, the study of different kind
of edges or defects in the edges are promising future research
topics. 

Recently, much experimental progress has been made in the realizations
of related models with ultra-cold bosonic gases.   
In particular, it is now possible to create
an artificial Kagome lattice for ultra-cold
bosons.\cite{kagome_ol} 
Moreover, a gas of Rydberg atoms was successfully loaded in
 a two-dimensional optical lattice, and the spatially ordered
structures induced by the short range repulsion interactions have been 
observed directly\cite{bloch1}. In principle it should be possible
to create sharp boundaries on such systems.\cite{bloch2}
Together with
recent advances in single-site addressability\cite{ad1,ad2,ad3},
the effects illustrated above therefore give a promising perspective of
actually taking pictures of strings and localized fractional
charges analogous to the snapshot in our simulations in
Fig.~\ref{fig3}.

\begin{acknowledgments}

We are thankful for useful discussions with 
Frank Pollmann, Immanuel Bloch, Ying Jiang, Xiaogang Wen and Jan
Zaanen. 
This work was supported by the "Allianz fuer Hochleistungsrechnen Rheinland-Pfalz"
and the DFG via the SFB/Transregio 49.

\end{acknowledgments}

%\begin{acknowledgments}
%
%We acknowledge Frank Pollmann, Ying Jiang, Liang He, Shijie Hu, Xiaogang Wen and Jan
%Zaanen for helpful discussions. This work is supported by the DFG
%via the SFB/Transregio 49.
%
%\end{acknowledgments}
% ----------------------------------------------------------------
\bibliographystyle{apsrev}
\end{document}